\documentclass[a4paper,10pt,twoside]{cpc-hepnp}

\usepackage{multicol}
\usepackage{graphicx}
\usepackage{graphics}
\usepackage{booktabs}
\usepackage{amssymb,bm,mathrsfs,bbm,amscd}
\usepackage[tbtags]{amsmath}
\usepackage{lastpage}
\usepackage{CJK}

\begin{document}

\fancyhead[c]{\small Chinese Physics C~~~Vol. 37, No. 1 (2013)
010201} \fancyfoot[C]{\small 010201-\thepage}
\footnotetext[0]{Received April 2013}

\title{Online charge calibration of LHAASO-WCDA---a study with the engineering array}


\author{%
\quad GAO Bo$^{1}$\email{gaobo@mail.ihep.ac.cn}%
\quad CHEN Ming-Jun$^{1}$
\quad GU Ming-Hao$^{1}$
\\\quad HAO Xin-Jun$^{2}$
\quad LI Hui-Cai$^{3}$
\quad WU Han-Rong$^{1}$
\\\quad YAO Zhi-Guo$^{1}$
\quad YOU Xiao-Hao$^{4}$
\quad ZHOU Bin$^{1}$ for
the LHAASO collaboration } \maketitle

\address{%
$^1$ Institute of High Energy Physics, Chinese Academy of Sciences, Beijing 100049, China\\
$^2$ University of Science \& Technology of China, Hefei 230026, China\\
$^3$ School of Physical Science and Technology, Southwest Jiaotong University, Chengdu, 610031, China\\
$^4$ Normal University of Hebei, Shijiazhuang, 050016, China\\}

\begin{abstract}
LHAASO-WCDA is a large ground-based water Cherenkov detector array planned to be built at Shangri-La, Yunnan Province, China. As a major component of LHAASO project, the main purpose of LHAASO-WCDA is to survey the northern sky for very-high-energy (above 100~GeV) gamma ray sources. To gain full knowledge of water Cherenkov technique and to well investigate engineering issues, a 9-cell detector array has been built at Yang-Ba-Jing site, neighboring to the ARGO-YBJ experiment. With the array, charge calibration methods for low and high ranges of the PMT readout are studied, whose result shows that a very high precision at several percentages can be reached. These calibration methods are proposed to be applied in the future LHAASO-WCDA project.
\end{abstract}

\begin{keyword}
LHAASO-WCDA, water Cherenkov, charge calibration
\end{keyword}

\begin{pacs}
98.70.Sa; 95.55.Vj; 29.40.Vj; 29.40.Mc
\end{pacs}

\begin{multicols}{2}

\section{Introduction}


In gamma astronomy, water Cherenkov is proved to be the most sensitive detection technique among all kinds of ground particle detector arrays, such as those made up of plastic scintillators and resistive plate chambers, mainly due to its amazing performance in background rejection power~\cite{byrum,smith}. The Milagro experiment~\cite{milagro} pioneered this technique, and next generation facilities like HAWC~\cite{hawc} and LHAASO-WCDA~\cite{caoz} that adopt this technique will be able to achieve the sensitivity more than 15 times better.\par

Spectral measurement of gamma ray emissions is one of the essential objectives of an observation.
Dominated by air shower fluctuations in the uncertainty of energy measurement, ground particle detector arrays usually show modest performance in this aspect. However, when the statistics reaches a considerable level, the only uncertainty remained comes from the detector itself, as the average behavior of the shower cascade can be modelled quite well by some code such as CORSIKA~\cite{corsika}. In this case, once a power law or similar spectrum were presumed, with the help of comparison to simulations, spectral and flux parameters would be able to be determined in a good precision, provided the error of the detector response simulation is well controlled. This therefore requires a good charge calibration of the detector, 5\% precision is good enough for ground-based detector array.\par

Calibrating the charge is usually not an easy task for air shower detectors, which are huge in size and frequently suffer from the variation of the environmental conditions. Artificial sources such as accelerator beam, radioactive source or light-emitting diode light are hard to control, and distributing them to all detector units time by time is someway difficult. The natural source like cosmic ray is the best for this purpose, but the application of it to a certain detector should be specifically studied on base of the precision requirement of the experiment and frequency request from the environmental variations.\par

Cosmic muons are commonly used for the charge calibration, as they are bombarding detectors all the time with a rather high rate and the interaction mechanism of it passing through the detector is well known and rather simple. In order to achieve such a goal, thin plastic scintillator detector of experiments such as AS$\gamma$ measures the signals of minimum ionizing particles~\cite{asgamma}, and water tank detector of experiments like Auger measures the vertical equivalent muon (VEM) signals~\cite{augervem}. As to water Cherenkov detector arrays such as LHAASO-WCDA, cosmic muons can be applied too, as they can form a special peak in the charge distribution~\cite{wcdproto}. One needs to investigate how to sort out the peak with a simpler setup or even without any additional auxiliary instrumentations. This is one of the goals of the study.\par

Besides muons, other cosmic secondary particles such as electrons, gammas may have a chance to contribute to the calibration too. This is not well established for air shower experiments, because the amplitude of the detector response generally depends on the varying energy of these particles. But for water Cherenkov detector array, the story may be different, as the signal of the single photoelectron dominates in the charge distribution. This is another concern of the work.\par

In this paper, first the LHAASO-WCDA engineering array is introduced, including key functional sub-systems; then the charge calibration methods are studied and presented in details, demonstrated with the measurement results.

\section{WCDA engineering array}
LHAASO-WCDA, a water Cherenkov detector array with an area of 90,000~$\rm{m}^{2}$, is planned to be built at Shangri-La, Yunnan Province, 4300 a.s.l., in next a few years. To gain full knowledge of the water Cherenkov technique and to well investigate the engineering issues, acting as a sequel of the prototype detector~\cite{wcdproto}, an engineering array of LHAASO-WCDA was built at Yang-Ba-Jing in 2010 and has been operated since then for more than 2 years.
\subsection{Water pool}
The engineering array of the water Cherenkov detector is located around 15~m northwest of the ARGO-YBJ experiment at an altitude of 4300~m a.s.l. The main part of the engineering array is a pool of water. The effective dimension of the pool is 15~m$\times$15~m at the bottom, with the pool walls concreted upward along a slope of 45$^{\circ}$ until 5~m in height, leading to 25~m$\times$25~m at the top. The roof of the pool is made of thin steel color-bond and insulated polyfoam, shielding the pool from external light and the nature. The whole pool is partitioned by black curtains into 3$\times$3 cells, each of which is 5~m$\times$5~m in size. A PMT is deployed at the bottom-center of each cell, facing upwards to collect the Cherenkov lights generated by air shower particles in water. Two manually movable shielding pads in size 1~m$\times$1~m are instrumented 15~cm above the photocathodes of two of these PMTs, for purpose of better discriminating the muon signals. To keep the water clean, a facility for purifying the water is built beside the pool, recirculating the pool water via pipe nets stretched into the pool.

\subsection{PMT}
Two kinds of 8-inch PMTs are deployed in the array: eight of type R5912 from Hamamatsu, and one of type 9354KB from ET Enterprises. A tapered voltage divider supplied with positive high voltage (HV) is adopted in the base circuit, which is potted to be water-proof with a special craft. With this base design and the readout from the anode, a dynamic range from 1/4 photoelectron (PE) to more than 700 PEs within a linearity level of 5\%, at the operating gain around 2$\times$10$^{6}$, is achieved. The PMT signal is transmitted to a preamplifier above the water surface via an 11~m cable, split into two signals, amplified to 25$\times$ and 1$\times$ individually. See figure~\ref{fig:amp} for details. The signals of these two amplifications, namely high gain and low gain, corresponding to low range and high range of the PMT signal, are finally delivered to the electronics in the control room on the bank via two 100~m cables, respectively.

\begin{center}
\includegraphics[width=0.8\linewidth]{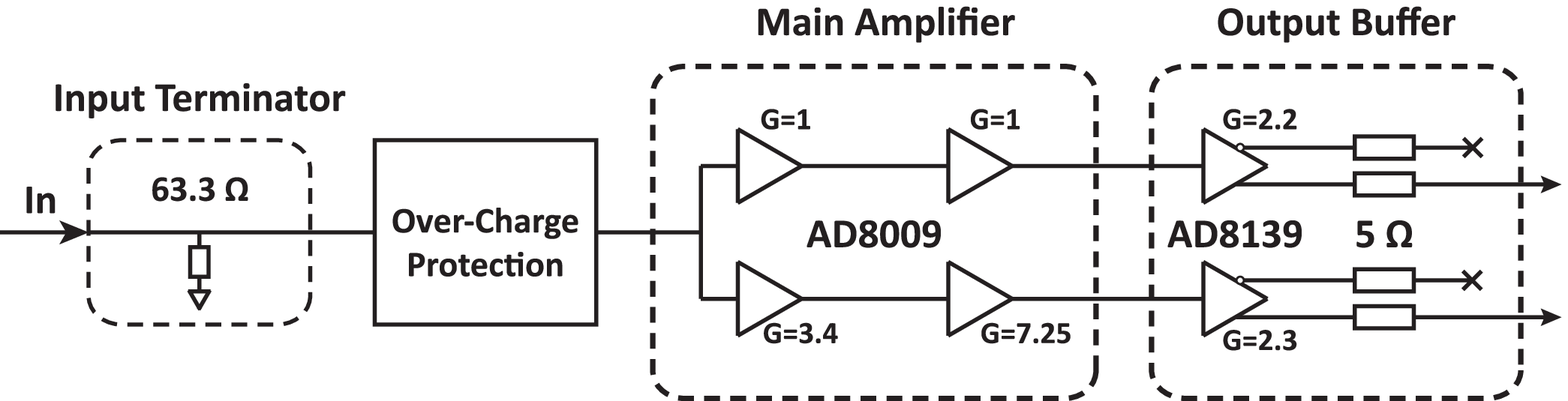}
\figcaption{\label{fig:amp}Diagram of the preamplifier. The signal from a PMT is split into two signals which are then amplified separately. The amplification factors are around 25$\times$ and 1$\times$ for the high and low gains, respectively.}
\end{center}

\subsection{Electronics \& DAQ}
A 9-U VME board with 9 pairs of channels is designed as the front-end electronics (FEE) to read out and digitalize the PMT signals. Each pair of channels process a pair of signals from a same PMT. The pair of signals are shaped, digitalized with two Analog-to-Digit converters (ADCs), and passed to a FPGA. Before shaping, the low range signal is forked and handed over to a discriminator. If the amplitude exceeds a configurable threshold, arriving time of the signal is digitalized with a Time-to-Digit converter (TDC), and at the same time the FPGA is signaled to start the peak finding algorithm to calculate the charge of the two signals. A time measurement and two charge measurements form a hit datum of the PMT.\par

The FEE integrates also the functions like GPS timing, clocking, triggering, data buffering, etc.
Versatile configurable trigger logics such as the case of 1 PMT fired, the case of several PMTs fired within a time window of 100~ns, are firmed in the FPGA ROM, to perform the event triggering. Controlled by a data acquisition (DAQ) system on a PowerPC module plugged in the same VME crate, the event stream consisting of hit data is sent to a desktop computer with the TCP/IP protocol via a network cable. Data are stored in a disk buffer of the computer, waiting for transferring and further offline processing.\par

To achieve multiple goals of the engineering array, a particular data-taking process with rotational discrimination thresholds and trigger logics is carried out during nearly the whole operation period. The rotation, being realized through re-configuring the FPGAs when starting every run, includes at least the following 3 categories: 1) single channel mode, where all PMTs but 1 are masked off with very high thresholds, and a threshold around 1/3~PE (in amplitude) is applied to the remaining PMT. In this case, the trigger logic is set to any one PMT being fired; 2) high threshold mode, where all the PMTs are set to a high threshold around 10~PE, with trigger logic of any one PMT being fired; 3) physics mode, where all PMTs are set to a same threshold around 1/3 PE, with the major trigger logic of any 3~PMTs being fired within a time window of 100~ns. Besides the above 3 modes, there are still some other rotation cases, which are rarely used and not relative to this study.

\subsection{Environmental sensors}
Environmental conditions of the control room and the water are measured once every several seconds by sensors connected to a slow control system. These measured parameters include temperatures of the control room, the outdoor, the water top and the water bottom, the pressures of the air and the water bottom. The water quality in the pool is automatically checked too a few times per day by a dedicated tube device for measuring the attenuation length of LED light in the water sampled from the pool. With the help of these sensors and tools, the experimental conditions are digitally recorded and well controlled.

\section{Charge calibration of the low range channel}
In the single channel mode of the data-taking, the PMT hits mostly comprise of single photoelectron (SPE) signals. These signals are mainly contributed from the low energy ($<$100~GeV) cosmic ray showers, whose major secondary components are photons, electrons (including positrons) and muons. The number of photons is about 10 and 20 times more than that of electrons and muons, respectively, at the altitude around 4300 m a.s.l., with energy greater than Cherenkov threshold in water. Photons interact with the water via Compton effect or pair production, yielding charged electron tracks and finally generating Cherenkov lights to be observed by PMTs. Taking the detection efficiency and energy distribution into account, the contribution ratio of photons, electrons and muons to the single counting rate is roughly estimated to be 4:1:1. Besides, when there are fresh water filling into the pool, the alpha decay of radioactive $Radon$ (${}^{222}Rn$) bringing in with the water produces fluorescence lights in the $H_{2}O$ molecular, radiating uniformly in direction and being collected by PMTs. Even when the pool is empty, radioactive Radon leaked from the bottom and bank of the pool diffuses inside the shut pool room, producing fluorescence lights in the air and generating PMT signals too. All above sources, working alone or together, lead to a high single counting rate of PMTs, and due to the low detection efficiency to these sources, SPE hits dominate in the PMT signals.\par
The charge distribution of low range output of a PMT is shown in figure~\ref{fig:spe} (top), where the bins around the peak position is fitted with a Gaussian function. Comparing it to the signal of very weak LED light dedicated for measuring the SPE curve, shown in the bottom of figure~\ref{fig:spe}, it is found that the peak position is almost same, indicating the peak of the single channel signals rides precisely at the SPE position.

\begin{center}
\includegraphics[width=0.65\linewidth]{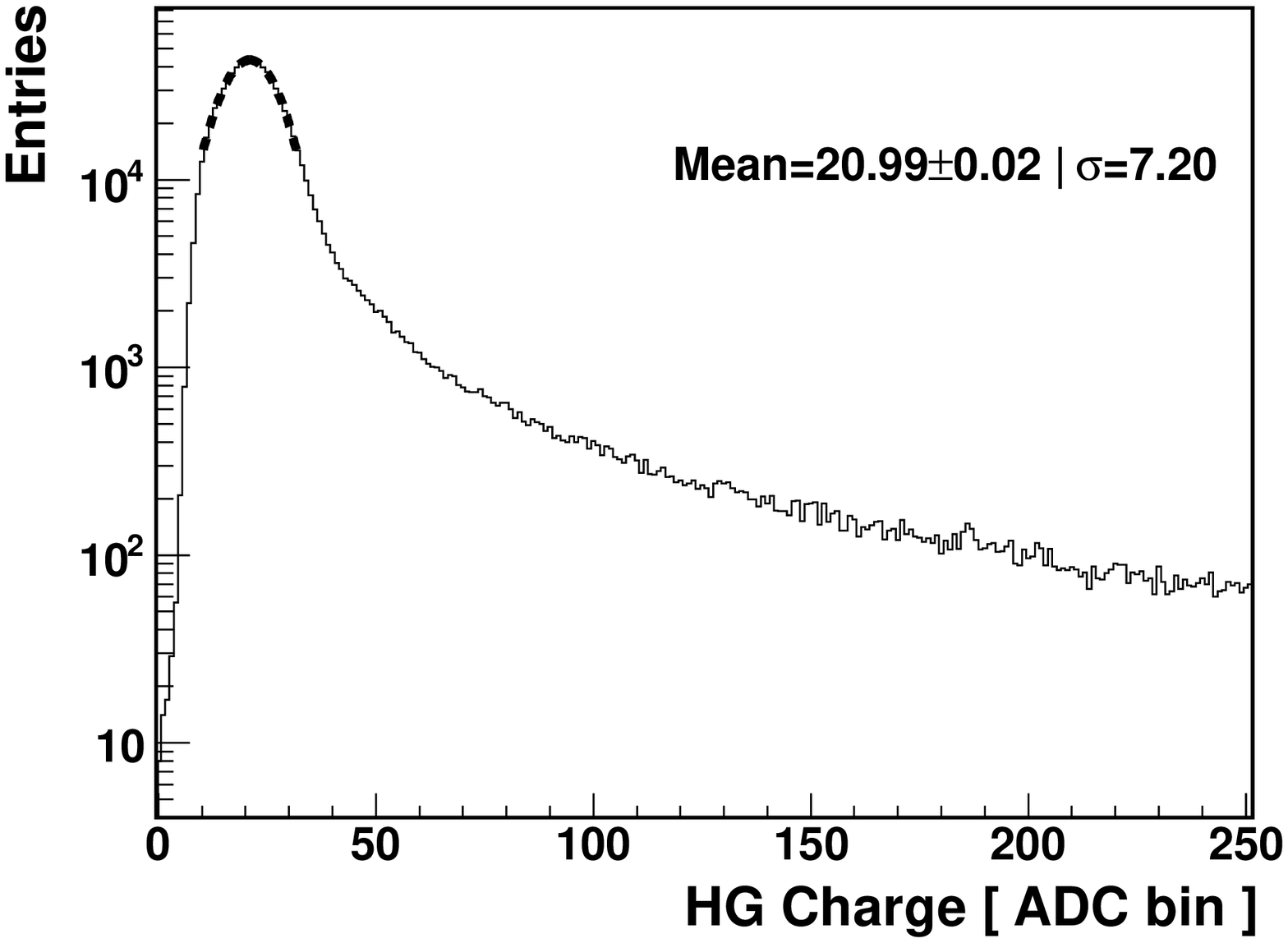}
\includegraphics[width=0.65\linewidth]{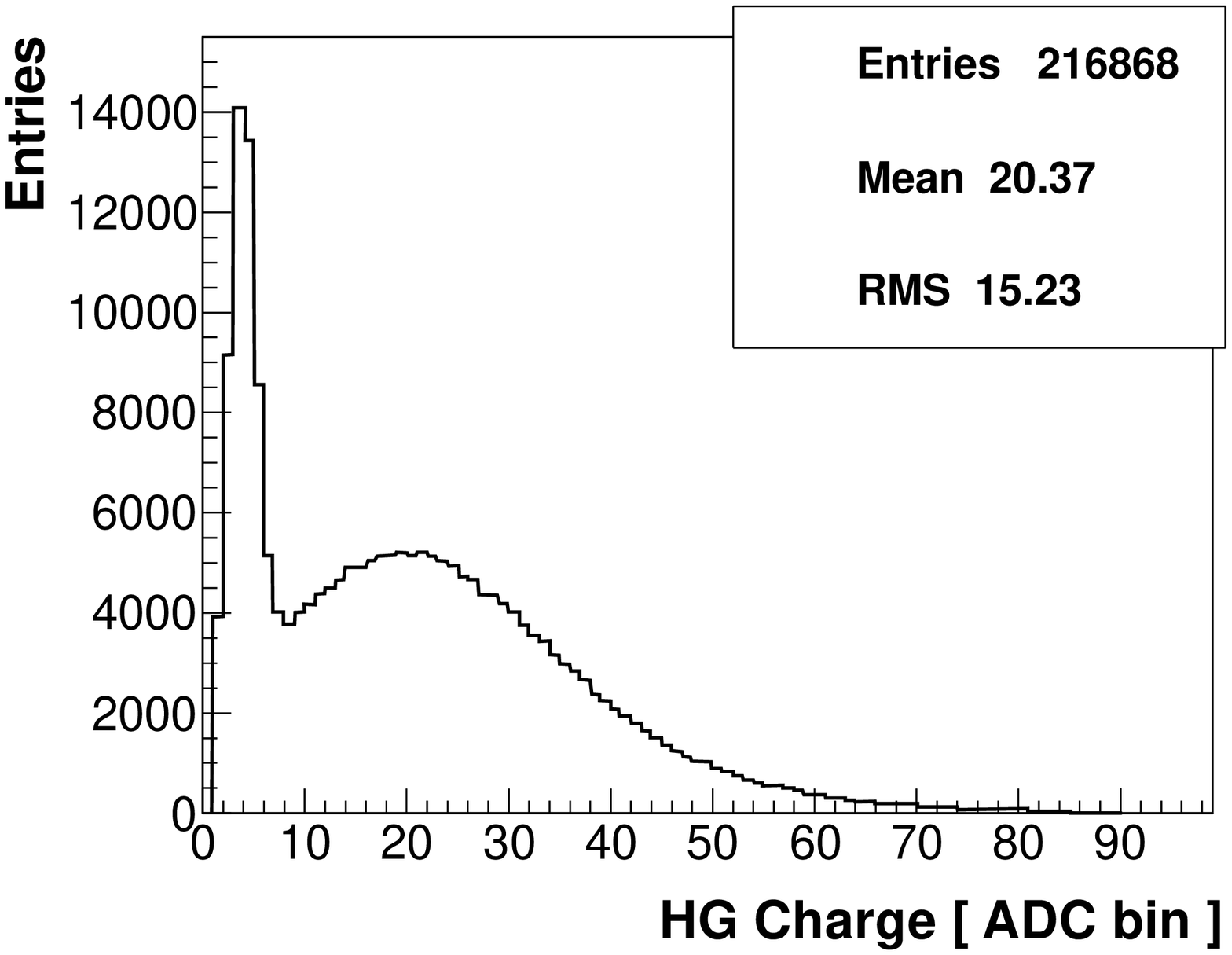}
\figcaption{\label{fig:spe}Top: charge distribution of low range output of a PMT, fitted with a Gaussian function; bottom: SPE curve obtained with an LED light source for the same PMT.}
\end{center}

The fact that the peak of single channel signals is the SPE peak can be demonstrated from another aspect. A global fitting to distribution in a wide range is made for the case when a shading pad is over the PMT top, as shown in figure~\ref{fig:spefit}, using a convolution function of a power law, a Poisson and a Gaussian, plus a single PE Gaussian, i.e.,

\begin{equation}\label{function:spefunc}
\footnotesize
  A\biggl[\int_{z_{1}}^{z_{2}}z^{\gamma}\sum_{n=1}^{N}\frac{e^{-z}z^{n}}{n!}\frac{1}{\sqrt{2{\pi}n}\sigma}e^{-\frac{(x-n{\mu})^{2}}{2n{\sigma}^2}}dz+\frac{B}{\sqrt{2\pi}\sigma}e^{-\frac{(x-\mu)^2}{2{\sigma}^2}}\biggr]
\end{equation}

Where the first term is for cosmic rays, and the second term is for noises like the dark noise and the radioactivity. The integral range, $z1$ and $z2$, is in the unit of ADC bin, starting from 0.01 PE up to at least 20\% more than the upper fitting range; the upper range of the summation $N$ is set to a value much bigger than the fitting range in the unit of PE. The other variables, such as $A$, $\mu$ , $\sigma$ , $\gamma$ and $B$, are parameters to be fit. For the case without the shading pad, it is very difficult to give a similar fit, as no regular function is found to well describe the muon components in the distribution.

\begin{center}
\includegraphics[width=0.7\linewidth]{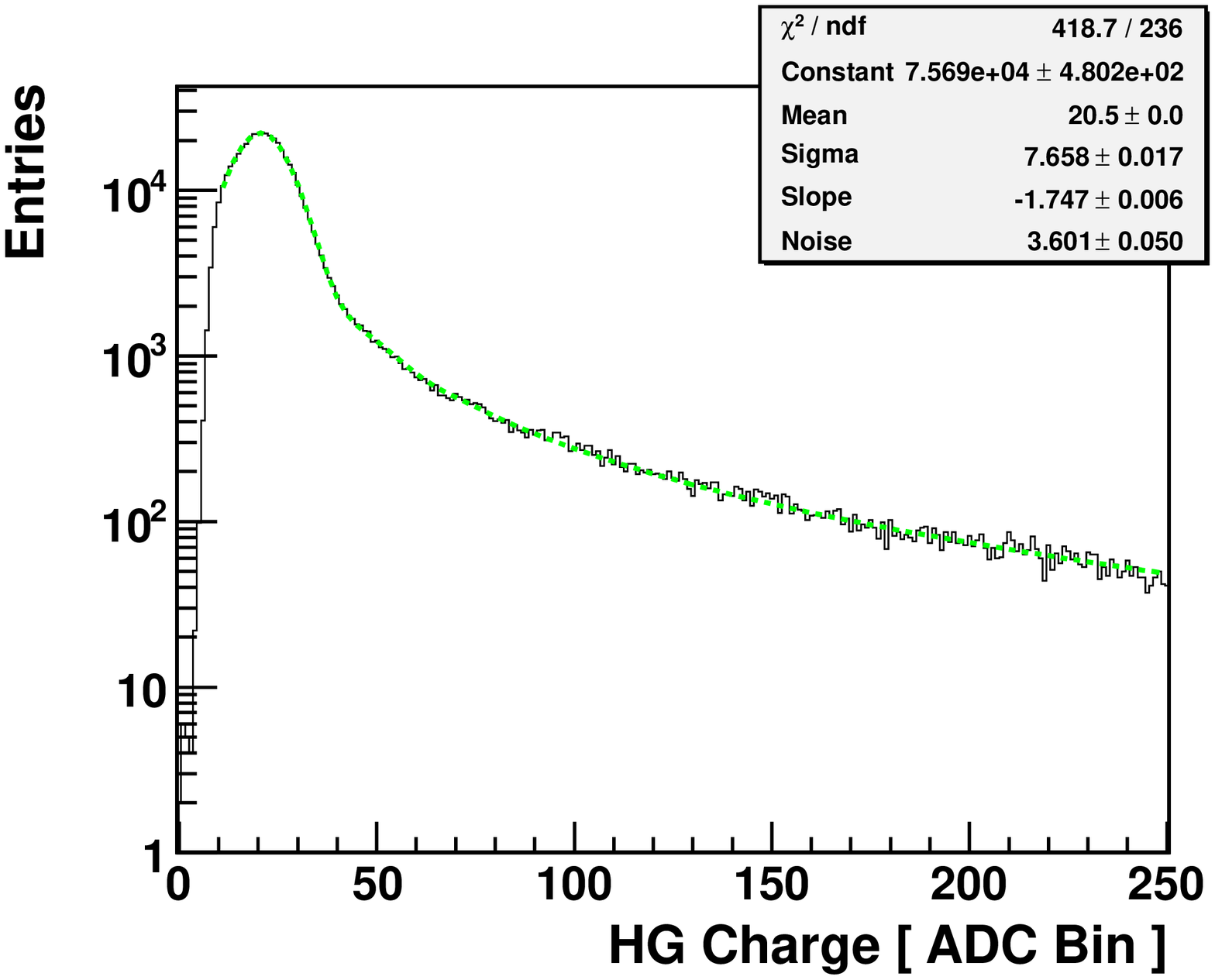}
\figcaption{\label{fig:spefit} Charge distribution of low range output of channel 7 PMT when a shading pad is on, fitted with a convolution function as equation 1. The names of the parameter are as the following: Constant = $A$, Mean = $\mu$, Sigma is equivalent to $\sigma$, Slope = $\gamma$, Noise = $B$.}
\end{center}

The SPE peaks of all PMTs are shown in table~\ref{tab:speall}, in which 30 seconds' data are taken for each PMT, all value are obtained from the charge distributions fitting by Gaussian function.
  \begin{center}
  \tabcaption{ \label{tab:speall} SPE peak value for all the PMTs.}
  \footnotesize
  \begin{tabular*}{80mm}{l@{\extracolsep{\fill}}cc}
    \toprule Channel No.  & Peak Value [ ADC bin ]  \\
    \hline
    1 (ET Tube) & 19.28$\pm$0.49 \\
    2 & 19.32$\pm$0.28 \\
    3 & 20.57$\pm$0.19 \\
    4 & 19.72$\pm$0.22 \\
    5 & 19.10$\pm$0.31 \\
    6 & 20.29$\pm$0.48 \\
    7 & 20.84$\pm$0.25 \\
    8 & 20.58$\pm$0.24  \\
    9 & 20.59$\pm$0.32  \\
  \bottomrule
\end{tabular*}
\end{center}

The SPE peak position represents an overall effect of the low range channel of the PMT, including the gain of the PMT, the amplification factor of the preamplifier, the attenuation of signals in the cable, the gain and the metric of the electronics. Especially the gain of the PMT, which is subjected to change during the operation in the first several years, can be monitored with the SPE peak. One should be aware that, the quantum efficiency and the collection efficiency of the PMT is not included in this kind of calibration.

\section{Charge calibration of the high range channel}
The previous prototype experiment~\cite{wcdproto} shows that near-vertical cosmic muons can produce a special peak at position around 350$-$600 PEs, relying on shape and type of the PMT. The peak is mainly formed by Cherenkov lights of muon tracks hitting the photocathode, so that very little dependence upon the water quality and depth is expected. With help of this feature, the PMT and high range channel of its electronics is foreseen to be calibrated, including overall effects such as the gain, the quantum efficiency and the collection efficiency of the PMT, the amplification factor of the preamplifier, the attenuation of signals in the cable, the gain and the metric of the electronics.\par

With the pool instead of tank configuration in the engineering array as well as the future full array,
coincident measurements with scintillators like the prototype experiment seem impractical. Monte Carlo simulations show that muons from all directions without the coincidence selection can form a peak too, but it is not very obvious, due to the disturbance from muons with large incident angles or far intersection points, and from high energy shower signals. Fortunately, proved by the simulation, a shading pad with a certain distance like 10$-$20~cm over the photocathode can well alleviate this disturbance and doesn't affect much the muon peak. That's why the devices of shading pads are put into two cells. Residing on rails, shading pads stay aside the PMTs in most occasions. To measure the muon peak, one can manually drag these pads along the rails, to ride over the top of the PMTs.\par

Nevertheless, the shading pad device, being a mechanical setup, is difficult to build and operate.
Instrumenting all PMTs of future big array with this kind of device is not a good solution. With a multiplication of power law of the charge (x-axis) to the entries (y-axis) can make the muon peak much clear, see in figure~\ref{fig:mppl}, where the power law index of 2.5 is used. Fitting the curve in the nearby range with a Gaussian function, the peak position is then obtained. Same analysis to the simulation data shows that the transform with a power law multiplication shifts the peak position a little bit higher, but less than 2\%. Anyway this shifting effect is uniform for all PMTs with the similar charge distribution, bringing nothing biased.

\begin{center}
\includegraphics[width=0.65\linewidth]{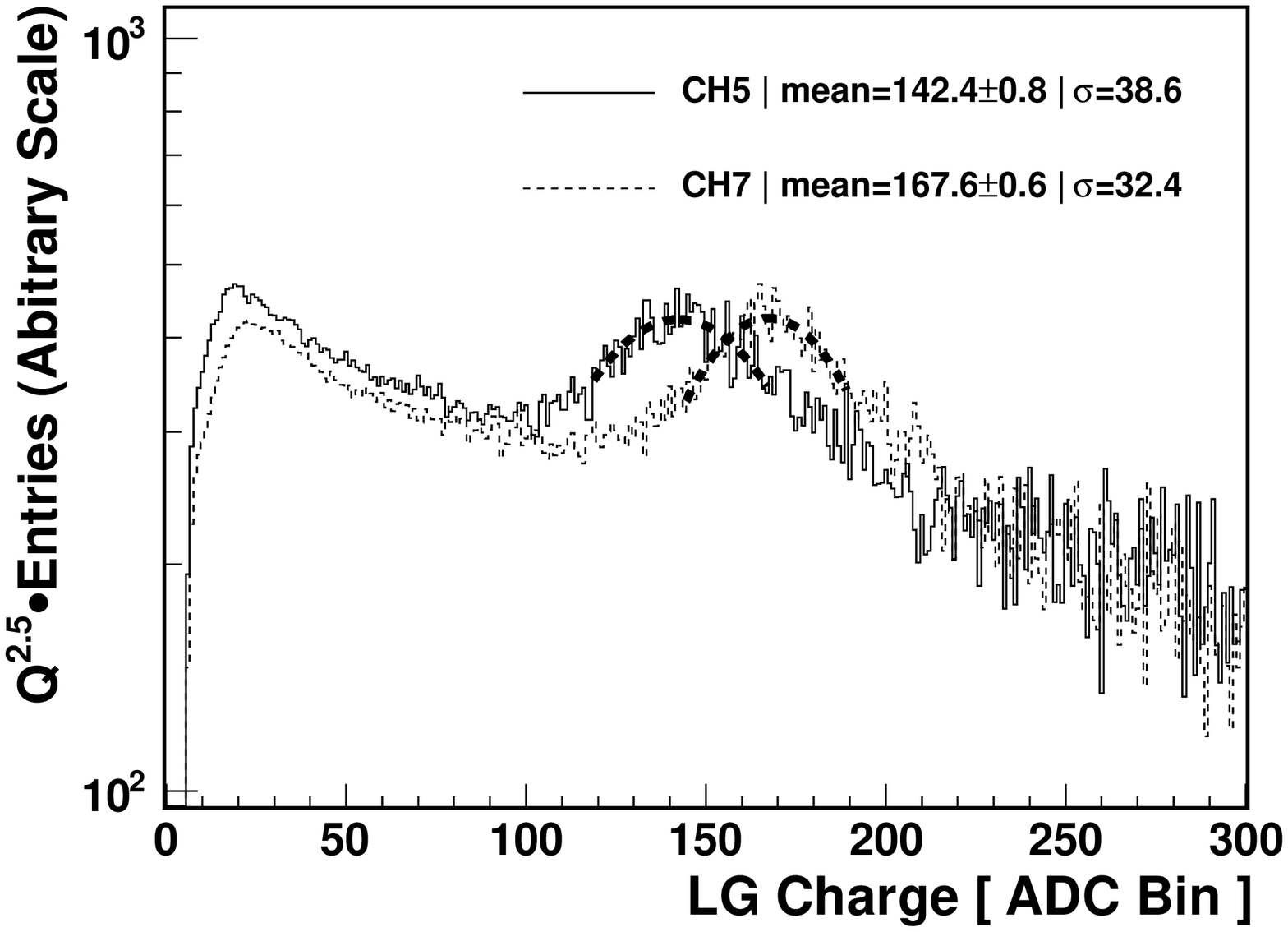}
\includegraphics[width=0.65\linewidth]{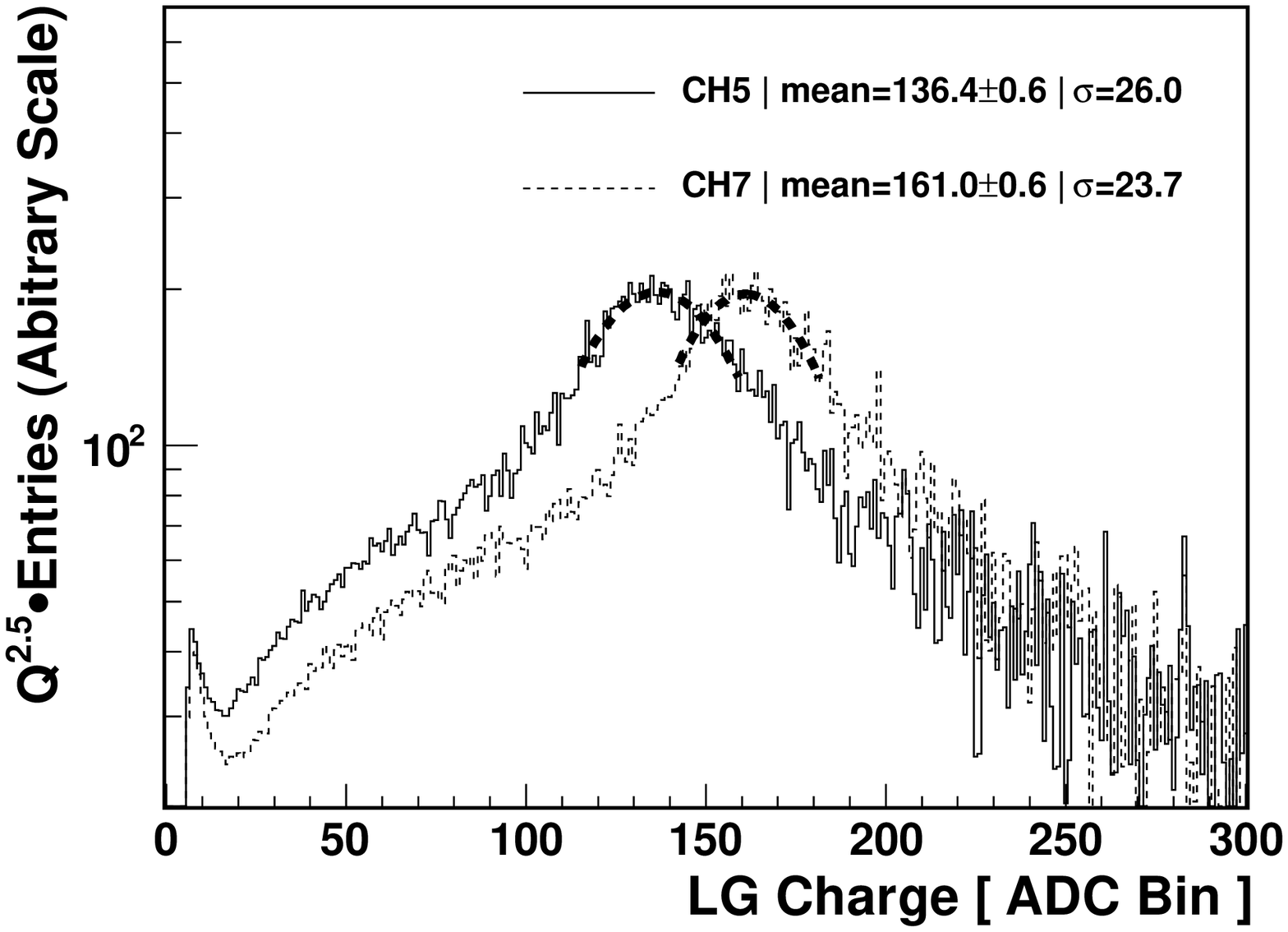}
\figcaption{\label{fig:mppl}Muon peaks of these two cells after a power law multiplication (power law index: 2.5), where Q is the number ADC bins divided by 150. Fitted with a Gaussian function, the peak position is obtained. Muon peak positions turn 4\% smaller in case of shading pads being on due to the factor of the $\delta-\rm{ray}$ signals generated in the muon tracks are screened by pads.}
\end{center}

With this kind of treatment, muon peaks of all PMTs are obtained almost effortlessly. The peak positions are not same, after  eliminating the PMT gain and  the front electronics' amplification between high and low range of each channel respectively, using one PMT as the reference,  the related differences between each channel are given, see in table~\ref{tab:mpdiff}, the result demonstrates almost pure difference between PMTs themselves. The ununiformity between Hamamatsu tubes is less than 10\%, but the difference between ET tube and other tubes is more than 20\%, such difference is mainly caused in two aspects: 1) the PMT shape differency,; 2) the quantum efficiency and the collection efficiency differences~\cite{wcdproto}.
  \begin{center}
  \tabcaption{ \label{tab:mpdiff} Muon peak value ratio between reference PMT and others.}
  \footnotesize
  \begin{tabular*}{80mm}{l@{\extracolsep{\fill}}ccc}
    \toprule Channel No.  & Ratio  \\
    \hline
    1 (ET Tube) & 0.734 \\
    2 & 0.967 \\
    3 & 1.003 \\
    4 & 1.058\\
    5 (Ref. PMT)& 1.0  \\
    6 & 1.093 \\
    7 & 1.085\\
    8 & 1.002  \\
    9 & 0.951  \\
  \bottomrule
\end{tabular*}
\end{center}

\section{Calibration stability}
To investigate the calibration stability and the dependence on environmental conditions, both low range and high range calibration are done intermittently (see section 2.3). The SPE peak and muon peak are fitted for every 30 seconds' and 30 minutes' data respectively. figure~\ref{fig:spetime} and figure~\ref{fig:mutime} shows the peak values distribution of one channel as the function of time in one particular month. During this month, the water depth and water quality changes quite a lot, and the temperature in the control room and in the water varies day by day. The SPE peak and muon peak position is quite stable, the variation evaluated by the average RMS of all channels is in the level around 2\%. That means a precision of 2\% of both low range charge and high range charge calibration can be achieved with this method, including all the environment effect factors to the PMT, cable and electronics of all channels.\\

\begin{center}
\includegraphics[width=0.8\linewidth]{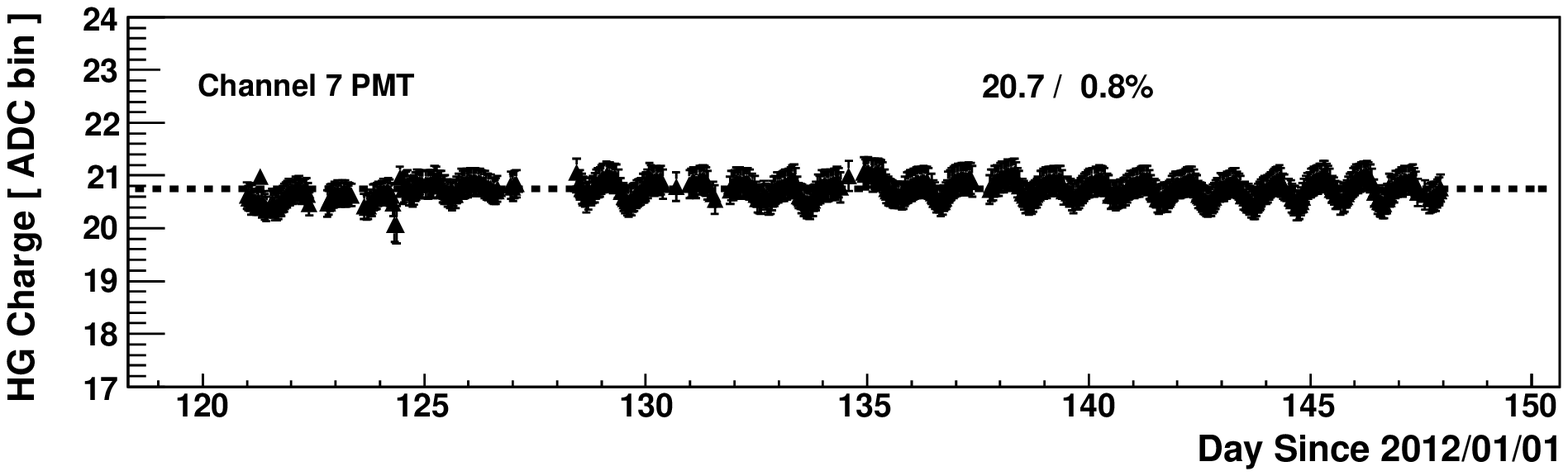}\\
\figcaption{\label{fig:spetime}SPE peak variation in one month. Fitted with a horizontal line.}
\end{center}
\begin{center}
\includegraphics[width=0.8\linewidth]{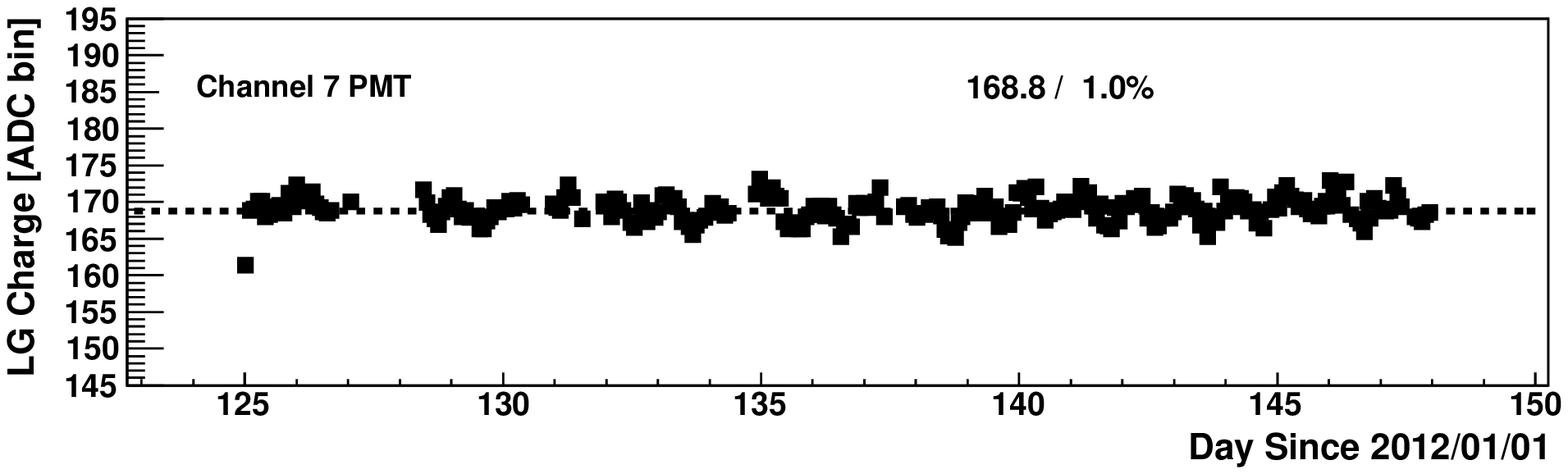}\\
\figcaption{\label{fig:mutime}Muon peak variation in one month. Fitted with a horizontal line.}
\end{center}

In these distributions, the daily variation of the peak position is observed. If drawing the peak position as the function of the temperature of the control room, a correlation around 0.2\%/$\rm{^\circ{C}}$ in average is found. This correlation can
be overall temperature effect, including the cables, most of which in the control room, and the PMTs, which reside in water, whose temperature is much more stable but still correlates with the room temperature. This temperature effect, which can be eventually corrected, is actually trivial in this analysis, as the data points used here are more or less concentrated in a small temperature range. Figure~\ref{fig:spetemp} and figure~\ref{fig:mutemp} show the correlation of SPE peak and muon peak value with room temperature of all PMTs mentioned above.

\begin{center}
\includegraphics[width=0.8\linewidth]{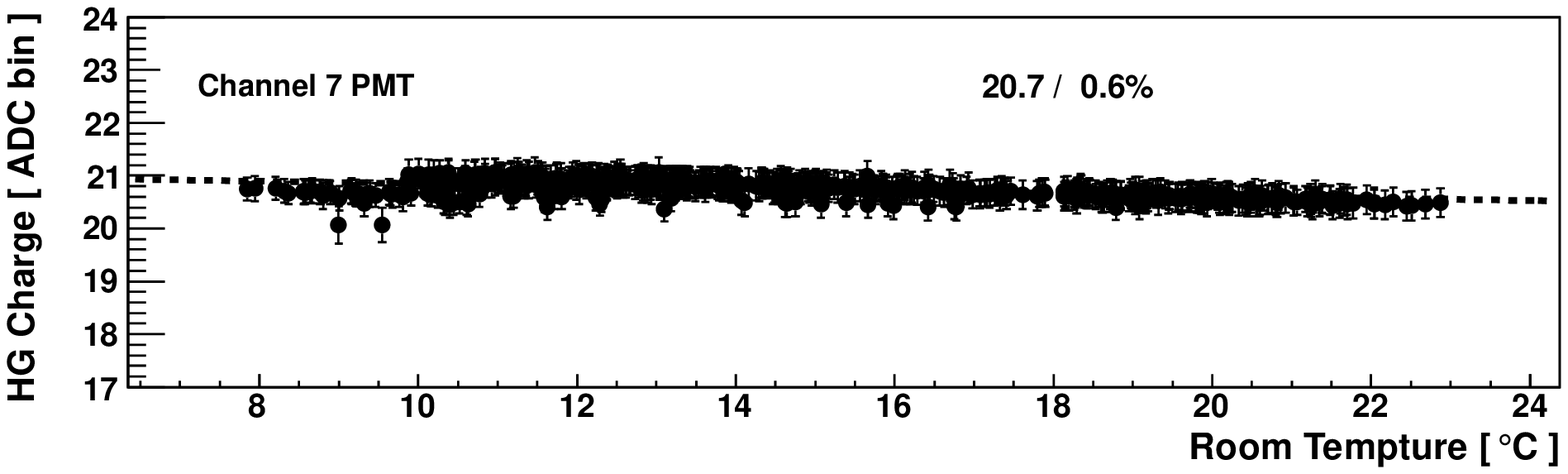}\\
\figcaption{\label{fig:spetemp}Correlation between SPE peak position and room temperature over one month. The RMS is the deviation of data points from the fitted line.}
\end{center}
\begin{center}
\includegraphics[width=0.8\linewidth]{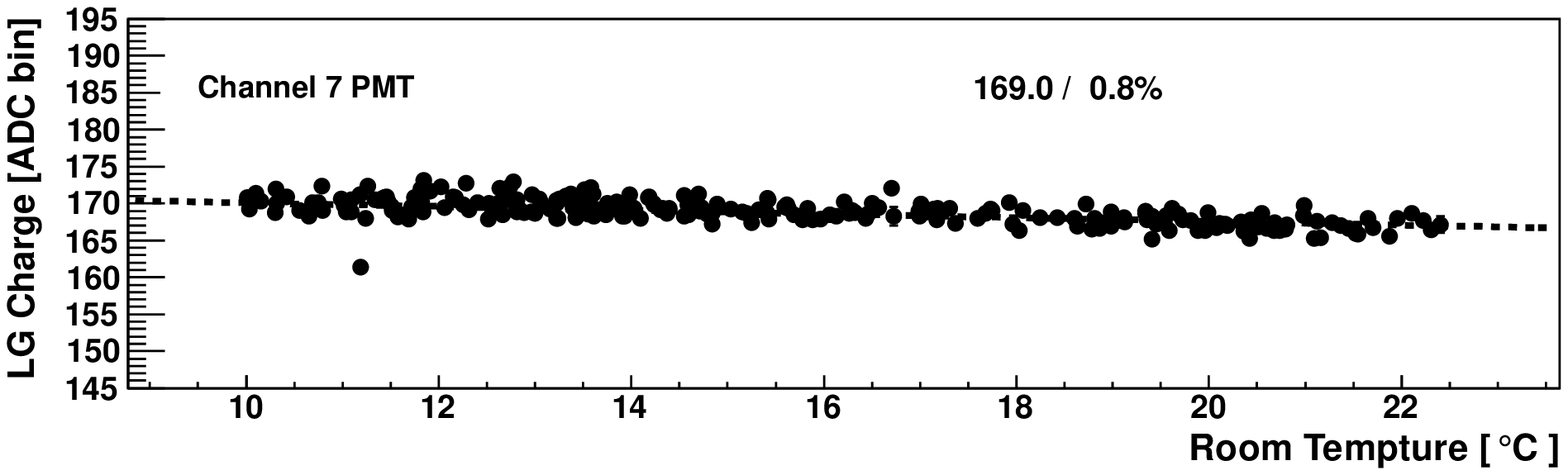}\\
\figcaption{\label{fig:mutemp}Correlation between muon peak position and room temperature in one month. The RMS is the deviation of data points from the fitted line.}
\end{center}

\section{Considerations of online charge calibration for LHAASO-WCDA}
Both the charges of high and low ranges can be well measured for known intensities by analyzing the single channel signals. In the future LHAASO-WCDA experiment, signals will be read out via two dynodes or a dynode and an anode in order to gain a wider dynamic range, and a triggerless mechanism for the data-taking will be adopted. Single channel signals of PMTs are digitalized by FEEs and transferred to a DAQ system composed of computer clusters, for a soft triggering and event building. At the same time of processing the hits data stream, the histograms for charge distribution of each PMT with both the low and the high range channels, as well as their correlations for the overlapped ranges, can be filled. For each PMT, analyzing online these 3 kinds of histograms, assuming very good linearity of the electronics, the muon peak position, the SPE peak position and the two channels' correlation coefficient can be got via simple fittings.
Expressed in formulae, they are:

\begin{equation}\label{function:q_hmu}
  N_{\rm{C},\rm{\mu},\rm{S}}Q\eta{G_{\rm{H}}}\beta_{\rm{H}}\mu_{\rm{H}}m_{\rm{H}} = q_{\rm{H},\rm{\mu}}
\end{equation}

\begin{equation}\label{function:q_lspe}
  G_{\rm{L}}\beta_{\rm{L}}\mu_{\rm{L}}m_{\rm{L}} = q_{\rm{L},\rm{SPE}}
\end{equation}

\begin{equation}\label{function:alpha}
  \frac{G_{\rm{H}}\beta_{\rm{H}}\mu_{\rm{H}}m_{\rm{H}}}{G_{\rm{L}}\beta_{\rm{L}}\mu_{\rm{L}}m_{\rm{L}}} = \alpha
\end{equation}

Where $N_{\rm{C},\rm{\mu},\rm{S}}$ is the known muon peak position in the unit of number of Cherenkov lights hitting the PMT cathode, obtained from the simulation and careful PMT tests, related with the shape and effective area of photo-cathode; $Q$ and $\eta$ is the quantum efficiency and the collection efficiency of the PMT respectively; $G_{\rm{H}}$ and $G_{\rm{L}}$ is the gain of the PMT for the high and low range respectively; $\beta_{\rm{H}}$ and $\beta_{\rm{L}}$ is the amplification factor of the preamplifier for the high and low range channel respectively; $\mu_{\rm{H}}$ and $\mu_{\rm{L}}$ is the attenuation coefficient of the cable for the high and low range channel respectively; $m_{\rm{H}}$ and $m_{\rm{L}}$ is the metric coefficient of FEE for the high and low range respectively; $q_{\rm{H},\rm{\mu}}$ is the measured muon peak position in the unit of ADC bin of the high range; $q_{\rm{L},\rm{SPE}}$ is the measured SPE peak position in the unit of ADC bin of the low range; $\alpha$ is the measured correlation coefficient of the overlapped signals of the high gain and the low gain channels. With equations 2 and 4, the overall calibration parameter of the low range channel,

\begin{equation}\label{function:speallpar}
  Q\eta{G_{\rm{L}}}\beta_{\rm{L}}\mu_{\rm{L}}m_{\rm{L}} = \frac{q_{\rm{H},\rm{\mu}}}{\alpha{N_{\rm{C},\rm{\mu},\rm{S}}}}
\end{equation}

is secured. And with these equations 2-4, the PMT performance, e.g., the product of $Q$ and $\eta$ is easy to be calculated, that is

\begin{equation}\label{function:qeta}
  Q{\eta} = \frac{1}{\alpha{N_{\rm{C},\rm{\mu},\rm{S}}}}\frac{q_{\rm{H},\rm{\mu}}}{q_{\rm{L},\rm{SPE}}}
\end{equation}

This parameter can act as a physical input for the simulation code, for a purpose of better understanding the detectors dynamically.

\section{Conclusion}
With the study carried out on the LHAASO-WCDA engineering array, a method for calibrating the charges for both the low and high range channels is developed. Natural sources of cosmic rays such as muons and photons can produce two kinds of nice peaks on PMTs, which can be nicely fitted with simple Gaussian functions. These peak positions are very stable, relying only on the PMT and the electronics, guaranteeing a precise calibration at the level 2\% can be achieved. These peaks can be obtained by online analyzing the single channel signals during a very short data-taking time window such as 30 seconds and 30 minutes, ensuring a PMT monitoring and a real-time calibration can be proceeded for the future LHAASO-WCDA experiment.

\section{Acknowledgement}
The authors would like to express their gratitude to X.F. Yuan, G. Yang, W.Y. Chen and C.Y. Zhao for their essential support in the installation, commissioning and maintenance of the engineering array.\\

\end{multicols}

\clearpage

\end{document}